# Yukawa potential for realistic prediction of Hubbard and Hund interaction parameters for transition metals[★]

Adam B. Cahaya[a,∗,1], Anugrah Azhar[b] and Muhammad Aziz Majidi[a]

[a]*Department of Physics, Faculty of Mathematics and Natural Sciences, Universitas Indonesia, Depok 16424, Indonesia, (62)21-7872610*
[b]*Physics Study Program, Faculty of Sciences and Technology, Syarif Hidayatullah State Islamic University Jakarta, Tangerang Selatan 15412, Indonesia, (62)21-7401925*

ABSTRACT

The generalized Hubbard model is an important theoretical model for modeling strongly correlated materials. To be able to theoretically predict the properties of the materials, the model requires Hubbard's $U$ and Hund's $J$ parameter that represent the on-site Coulomb and exchange interaction, respectively. For bare Coulomb interactions, the analytic expression of $U$ and $J$ are analytically described by Racah's $A, B, C$ parameters. However, the values of $U$ are too large for transition-metal-based materials. To obtain more accurate values for realistic materials, we employ a Yukawa-type screened Coulomb interaction, characterized by screening constant $\lambda$. We characterize $\lambda$ for transition metals. The modified $A, B, C$ parameters give a more realistic $U$ and $J$ values.

## 1. Introduction

Hubbard model and its variants are very useful tools in the theory of condensed matter systems[1]. Its utilization in the meanfield theory has been shown to give realistic descriptions of materials [2]. Key ingredients for utilizing a generalized Hubbard model to study a magnetic material are Hubbard and Hund parameters[3, 4]. Hubbard parameter $U$ and Hund parameter $J$ represent the on-site Coulomb and exchange interaction, respectively. The latter interaction is important in modeling the magnetism of strongly correlated materials, such as transition metals[5].

While the Hubbard model is an approximate model, $U$ and $J$ values can be analytically described in terms of Racah's three parameters: $A, B, C$, which is related to Slater integrals involving bare Coulomb interaction. Bare Coulomb interaction gives a large and unrealistic value of $U$. Density functional theory (DFT) studies employ a screening mechanism to reduce the Coulomb interaction, which in turn give a more realistic value [6, 7]. However, an accurate understanding of the electronics properties of strongly correlated materials requires a better understanding of the screened Coulomb interaction [7]. Furthermore, Racah parameters suggests that the value of $J$ depends on the orbital pairs. Because of that, since Hubbard and Hund parameters are related through Racah parameters, it give rises to an inconsistency.

We aim to solve the inconsistency by deriving the Racah's parameters for a screened Coulomb potential. For the screened potential, Yukawa-type potential appears as a natural alternative[8]. Recently, rigorous mathematical expression of the Slater integrals of Yukawa potential has been studied and utilized in chemical [8] and atomic [9, 10] physics. This article aims to show its application for determining the realistic parameters in material and condensed matter physics.

This manuscript is organized as follows. Relation of Hubbard and Hund parameters to Racah parameters is discussed in Sec 2. In Sec. 3.1, we derive expressions of the Slater integrals for screened Coulomb interaction that are more straight-forwardly comparable to those of bare Coulomb interaction. In Sec. 3.2 we show that the modified $A, B, C$ parameters satisfy the same relation as the bare Coulomb interaction. We also discuss the cases when $J$ is independent to the orbital pairs. Last but not least, we give the realistic Hubbard and Hund parameters in Sec. 4 and conclude our results in Sec. 5. We hope to give a guide for estimating Hubbard and Hund parameters for future studies.

## 2. Generalized Hubbard model and Racah parameters

Hubbard model is an important model in the theoretical study of strongly correlated materials. In Generalized Hubbard model, the leading term of the interaction Hamiltonian can be written as

$$H_{int} = U \sum_{i\alpha} n_{i\alpha\uparrow} n_{i\alpha\downarrow} + \sum_{i,\alpha<\beta} \left(U - \frac{3}{2} J_{\alpha\beta}\right) n_{i\alpha} n_{i\alpha} - 2 \sum_{i,\alpha<\beta} J_{\alpha\beta} \mathbf{S}_{i\alpha} \cdot \mathbf{S}_{i\beta} \quad (1)$$

where $U$ and $J_{\alpha\beta}$ are the on-site Coulomb and exchange interaction constants, respectively. As a comparion, a standard Hubbard model only focuses the on-site interaction. We expand the bare Coulomb potential $V(|\mathbf{r}_1 - \mathbf{r}_2|) = e^2/\left(4\pi\varepsilon_0 |\mathbf{r}_1 - \mathbf{r}_2|\right)$ with the following spherical harmonics expansion

$$\frac{1}{|\mathbf{r}_1 - \mathbf{r}_2|} = \sum_{l=0}^{\infty} \frac{1}{2l+1} \frac{r_<^l}{r_>^{l+1}} \sum_{m=-l}^{l} Y_{lm}(\mathbf{\Omega}_1) Y_{lm}^*(\mathbf{\Omega}_2), \quad (2)$$

where $r_< = \min(r_1, r_2)$, $r_> = \max(r_1, r_2)$, $Y_{lm}(\mathbf{\Omega})$ is the spherical harmonic function, $r = |\mathbf{r}|$ and $\mathbf{\Omega} \equiv \mathbf{r}/r$. The values of $U$ and $J_{\alpha\beta}$ can be written in terms of the Racah parameters $A, B$ and $C$

$$U = A + 4B + 3C. \quad (3)$$

The values of $J_{\alpha\beta}$ is listed in Table 1. Although $J_{\alpha\beta}$ depends on the orbital pairs, a model for strongly correlated materials usually employs a pair-independent value of $J_{\alpha\beta} = J$. This inconsistency is solved in Sec. 4.

[★]This document is the results of the research project funded by Kementerian Riset Teknologi Dan Pendidikan Tinggi Republik Indonesia through PDUPT Research Grant No. NKB.1602/UN2.R3.1/HKP05.00/2019.

✉ adam@sci.ui.ac.id (A.B. Cahaya); aziz.majidi@sci.ui.ac.id (M.A. Majidi)
ORCID(s): 0000-0002-2068-9613 (A.B. Cahaya); 0000-0002-0613-1293 (M.A. Majidi)

**Table 1**
Values of interorbital exchange elements $J_{\alpha\beta}$ for $3d$ orbitals in terms of the Racah parameters.

| orbital | $xy$ | $yz$ | $xz$ | $x^2-y^2$ | $3z^2-r^2$ |
|---|---|---|---|---|---|
| $xy$ | 0 | 3B+C | 3B+C | C | 4B+C |
| $yz$ | 3B+C | 0 | 3B+C | 3B+C | B+C |
| $xz$ | 3B+C | 3B+C | 0 | 3B+C | B+C |
| $x^2-y^2$ | C | 3B+C | 3B+C | 0 | 4B+C |
| $3z^2-r^2$ | 4B+C | B+C | B+C | 4B+C | 0 |

The Racah parameters are related to Slater integrals $F^l$

$$\begin{pmatrix} A \\ B \\ C \end{pmatrix} = \frac{e^2}{4\pi\epsilon_0} \begin{pmatrix} 1 & 0 & -\frac{1}{9} \\ 0 & \frac{1}{49} & -\frac{5}{441} \\ 0 & 0 & \frac{35}{441} \end{pmatrix} \begin{pmatrix} F^0 \\ F^2 \\ F^4 \end{pmatrix}, \quad (4)$$

$$F^l = \int_0^\infty r_1^2 dr_1 \int_0^\infty r_2^2 dr_2 R^2(r_1) R^2(r_2) \frac{r_<^l}{r_>^{l+1}}. \quad (5)$$

Here, $R(r)$ is the normalized radial part of electron orbital, $r_> = \max\{r_1, r_2\}$ and $r_< = \min\{r_1, r_2\}$.

For Slater-type $3d$-orbital $R(r) = \sqrt{8\zeta^7/45}\, r^2 e^{-\zeta r}$ under bare Coulomb interaction, the Slater integrals $F^l$ can be written as follows.

$$F^0 = \frac{793}{3072}\zeta \quad F^2 = \frac{2093}{15360}\zeta \quad F^4 = \frac{91}{1024}\zeta \quad (6)$$

## 3. Modified $A, B, C$ parameters for screened Coulomb potential

In transition-metal oxides, the attenuation of the Coulomb interaction $V(\mathbf{r})$ can be described by Thomas-Fermi screening [11]. Thomas-Fermi screening can be described in terms of screening function $\phi$ as follows.

$$V(\mathbf{r}) = \frac{e^2}{4\pi\epsilon_0 r}\phi(r), \quad (7)$$

where $\phi(r)$ is the solution of the following non-linear differential equation [12, 13]

$$\left[-\frac{\hbar^2}{2m}\nabla^2 - V(\mathbf{r})\right]\phi(r) = \phi(r). \quad (8)$$

In transition-metal oxides, the leading term of the correction can be written as follows [14]:

$$V(\mathbf{r}) = \frac{e^2}{4\pi\epsilon_0 r} e^{-\lambda r}. \quad (9)$$

Note that for large $\lambda$, the interaction become short-ranged and $V(\mathbf{r})$ become a delta function. Therefore, the screening of the bare Coulomb interaction of two charges at $\mathbf{r}_1$ and $\mathbf{r}_2$ manifests as a Yukawa potential-type screened Coulomb interaction.

$$V(|\mathbf{r}_1 - \mathbf{r}_2|) = \frac{e^2}{4\pi\epsilon_0 |\mathbf{r}_1 - \mathbf{r}_2|} e^{-\lambda|\mathbf{r}_1 - \mathbf{r}_2|}. \quad (10)$$

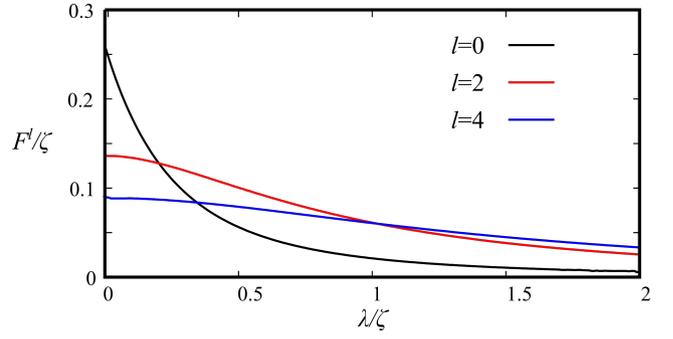

**Figure 1:** [Color online] Modified Slater integral $\tilde{F}^l$ of Yukawa-type screened Coulomb interaction $\frac{e^2}{4\pi\epsilon_0} e^{-\lambda|\mathbf{r}_1-\mathbf{r}_2|}/|\mathbf{r}_1 - \mathbf{r}_2|$ with screening constant $\lambda$. $\tilde{F}^0$ (black line), $\tilde{F}^2$ (red line) and $\tilde{F}^4$ (blue line) monotonically decrease as $\lambda$ increases. They determine the $A, B, C$ parameters, which in turn determine Hubbard's $U$ and Hund's $J$ parameters.

### 3.1. Slater integral for screened Coulomb potential

Before examining how $F^0$, $F^2$ and $F^4$ should be modified, we need to find the spherical harmonics expansion of Yukawa potential.

First we substitute the following plane wave expansion to the exponential functions of expression inside the integral in Eq. 10

$$e^{i\mathbf{q}\cdot\mathbf{r}} = 4\pi \sum_{l=0}^\infty i^l j_l(qr) \sum_{m=-l}^l Y_{lm}(\mathbf{\Omega}_q) Y_{lm}^*(\mathbf{\Omega}), \quad (11)$$

where $j_n(x)$ is the $n$-th spherical Bessel function. The integration over $d\mathbf{\Omega}_q$ can be simplified by spherical harmonics identities $\int d\mathbf{\Omega}_q Y_{l_1 m_1}^*(\mathbf{\Omega}_q) Y_{l_2 m_2}(\mathbf{\Omega}_q) = \delta_{l_1 l_2}\delta_{m_1 m_2}$, and taking integral over $q$. We then arrive at the spherical harmonics expansion of the Yukawa-type screened Coulomb potential

$$\frac{e^{-\lambda|\mathbf{r}_1-\mathbf{r}_2|}}{|\mathbf{r}_1 - \mathbf{r}_2|} = \lambda \sum_{l=0}^\infty i_l(\lambda r_<) k_l(\lambda r_>) \sum_{m=-l}^l Y_{lm}(\mathbf{\Omega}_1) Y_{lm}^*(\mathbf{\Omega}_2), \quad (12)$$

where $i_n(x)$ and $k_n(x)$ is the $n$-th modified spherical Bessel function of the first[15] and second[16] kind, respectively (see 3). Eq. 12 is equivalent with the expansion derived from Ref. [9, 10]. In Sec. 3.2 we show that asymptotic behavior of modified spherical Bessel functions can be used to simplify Eq. 12 for large $\lambda$. Therefore $r_<^l/r_>^{l+1}$ terms of bare-Coulomb interaction should be replaced by

$$f_l(r_1, r_2) = (2l+1)\lambda i_l(\lambda r_<) k_l(\lambda r_>) \quad (13)$$

$$= \begin{cases} \frac{r_<^l}{r_>^{l+1}}\left(1 - \delta_{l0}\lambda r_> + \frac{\lambda^2}{2}\left(\frac{r_<^2}{2l+3} - \frac{r_>^2}{2l-1}\right)\right), & \text{for } \lambda r_{1,2} \ll 1, \\ \frac{2l+1}{2}\frac{e^{-\lambda|r_1-r_2|}}{\lambda r_1 r_2}\left(1 - \frac{l(l+1)}{2\lambda}\left|\frac{1}{r_1} - \frac{1}{r_2}\right|\right), & \text{for } \lambda r_{1,2} \gg 1. \end{cases} \quad (14)$$

Here we write the asymptotic value for weak and strong screening limits, which can be obtained from the asymptotic behaviors of the modified spherical Bessel functions (see A). The modified Slater integral $\tilde{F}^k$ can be found by replacing $r_<^l/r_>^{l+1}$ in Eq. 5 with $f_l(r)$,

$$\tilde{F}^k = \int_0^\infty \int_0^\infty r_1^2 dr_1 r_2^2 dr_2 R^2(r_1) R^2(r_2) f_l(r_1, r_2). \quad (15)$$

Fig. 1 shows particular values of $\tilde{F}^0$, $\tilde{F}^2$ and $\tilde{F}^4$.

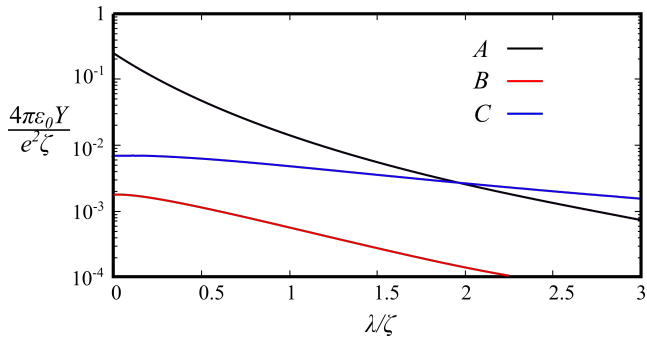

**Figure 2:** [Color online] $Y = \{A, B, C\}$, where $A$ (black line), $B$ (red line) and $C$ (blue line) parameters for Yukawa-type screened Coulomb interaction monotonically decrease as $\lambda/\zeta$ increases. For large $\lambda \gg \zeta$, $A$ and $B$ are order of magnitude smaller than $C$. Their asymptotic values of $A$, $B$ and $C$ are $\frac{e^2}{4\pi\epsilon_0} 7\zeta^5 \lambda^{-4}/72$, $\frac{e^2}{4\pi\epsilon_0} \zeta^5 \lambda^{-4}/144$ and $\frac{e^2}{4\pi\epsilon_0} 5\zeta^3 \lambda^{-2}/256$, respectively.

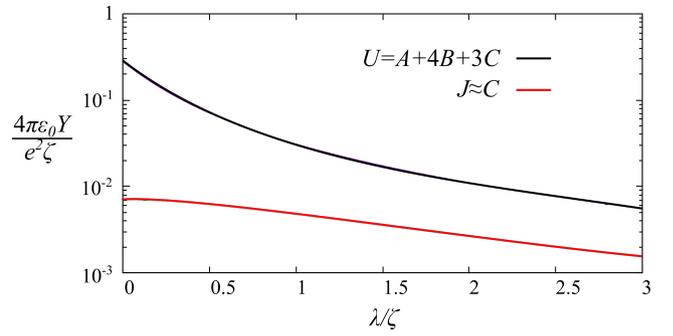

**Figure 3:** [Color online] $Y = \{U, J\}$, where $U$ (black line) and $J$ (red line) are Hubbard's and Hund's parameters for Yukawa-type screened Coulomb interaction, respectively. $U$ and $J$ monotonically decrease as $\lambda/\zeta$ increases. For small $\lambda \ll \zeta$, $U$ decreases faster than $J$. Their asymptotic values for large $\lambda \gg \zeta$ are $U \approx 3C$ and $J \approx C$, respectively.

### 3.2. Modified $A, B, C$ parameters

The $A, B, C$ parameters for the screened Coulomb interaction are

$$A = \frac{e^2}{4\pi\epsilon_0} \int_0^\infty r_1^2 dr_1 \int_0^\infty r_2^2 dr_2 R^2(r_1) R^2(r_2)$$
$$\times \lambda \left[ i_0(\lambda r_<) k_0(\lambda r_>) - i_4(\lambda r_<) k_4(\lambda r_>) \right], \quad (16)$$

$$B = \frac{e^2}{4\pi\epsilon_0} \int_0^\infty r_1^2 dr_1 \int_0^\infty r_2^2 dr_2 R^2(r_1) R^2(r_2)$$
$$\times \frac{5\lambda}{49} \left[ i_2(\lambda r_<) k_2(\lambda r_>) - i_4(\lambda r_<) k_4(\lambda r_>) \right], \quad (17)$$

$$C = \frac{e^2}{4\pi\epsilon_0} \int_0^\infty r_1^2 dr_1 \int_0^\infty r_2^2 dr_2 R^2(r_1) R^2(r_2)$$
$$\times \frac{5\lambda}{7} i_4(\lambda r_<) k_4(\lambda r_>), \quad (18)$$

**Weak screening limit-** From Eq. 14, one can see that as $\lambda$ goes to zero, $\lim_{\lambda \to 0} f_l(r_1, r_2) = r_<^l/r_>^{l+1}$ and $\lim_{\lambda \to 0} \tilde{F}^k = F^k$ reduce to that of the bare Coulomb interaction. Since first-order correction only appears for $l = 0$, $\tilde{F}^0$ is more sensitive to screening. $F^0$, $F^2$ and $F^4$ decrease as $\lambda$ increases. For small $\lambda$,

$$\lim_{\lambda \ll \zeta} \tilde{F}^0 = \frac{793}{3072}\zeta - \lambda + \frac{\lambda^2}{\zeta}\frac{90181}{36864} \quad (19)$$

$$\lim_{\lambda \ll \zeta} \tilde{F}^2 = \frac{2093}{15360}\zeta - \frac{\lambda^2}{\zeta}\frac{8983}{36864} \quad (20)$$

$$\lim_{\lambda \ll \zeta} \tilde{F}^4 = \frac{91}{1024}\zeta - \frac{\lambda^2}{\zeta}\frac{195}{4096} \quad (21)$$

The corresponding $A, B, C$ parameters are

$$\lim_{\lambda \ll \zeta} A = \frac{e^2}{4\pi\epsilon_0} \left( \frac{143}{576}\zeta - \lambda + \frac{\lambda^2}{\zeta}\frac{401323}{165888} \right) \quad (22)$$

$$\lim_{\lambda \ll \zeta} B = \frac{e^2}{4\pi\epsilon_0} \left( \frac{143}{80640}\zeta - \frac{\lambda^2}{\zeta}\frac{4979}{903168} \right) \quad (23)$$

$$\lim_{\lambda \ll \zeta} C = \frac{e^2}{4\pi\epsilon_0} \left( \frac{65}{9216}\zeta - \frac{\lambda^2}{\zeta}\frac{325}{86016} \right) \quad (24)$$

Therefore, the Hund's parameter $J$ decrease slower than the Hubbard's parameter $U = A + 4B + 3C$ This is because it depends only on the second order correction and higher.

**Strong screening limit-** Using $f_l(r_1, r_2)$ for large $\lambda$ (see Eq. 14), one can see that when $\lambda$ approaches infinity,

$$\lim_{\lambda \to \infty} \tilde{F}^0 : \tilde{F}^2 : \tilde{F}^4 \to 1 : 5 : 9. \quad (25)$$

Therefore $A \propto (F^0 - F^4/9)$ and $B \propto (F^2 - 5F^9/9)$ are orders of magnitude smaller than $C$ for large $\lambda \gg \zeta$. The limiting behavior of $A, B, C$ parameters (to their respective leading orders)

$$\lim_{\lambda \gg \zeta} A : B : C = \frac{14\zeta^2}{\lambda^2} : \frac{\zeta^2}{\lambda^2} : \frac{45}{16}, \quad (26)$$

where

$$\lim_{\lambda \gg \zeta} C = \frac{e^2}{4\pi\epsilon_0}\frac{5\zeta^3}{256\lambda^2}. \quad (27)$$

From the values of $A, B, C$ parameters, we can determine the Hubbard's $U$ and Hund's $J$ parameters. In the next section, we focus our study on the implication of the screened Coulomb interaction on the Hubbard's and Hund's parameters of transition-metal oxides.

### 4. Screening parameters for transition metals with realistic Hubbard's $U$ and Hund's $J$ parameters

Substituting Eqs. 16-18 to $U = 3A + 4B + C$, we arrive at the full expression of Hubbard parameter for screened Coulomb interaction

$$U = \frac{e^2}{4\pi\epsilon_0} \int_0^\infty r_1^2 dr_1 \int r_2^2 dr_2 R^2(r_1) R^2(r_2) \lambda \left[ i_0(\lambda r_<) k_0(\lambda r_>) \right.$$
$$\left. + \frac{20}{49} i_2(\lambda r_<) k_2(\lambda r_>) + \frac{36}{49} i_4(\lambda r_<) k_4(\lambda r_>) \right]. \quad (28)$$

$U$ decreases fast as $\lambda$ increases. The first order of $\lambda$ gives the following correction for $U$ for weak screening

$$\lim_{\lambda \ll \zeta} U \simeq U_{\text{bare Coulomb}} - \frac{e^2}{4\pi\epsilon_0}\lambda. \quad (29)$$

When the screening is very strong, $A$ and $B$ vanish faster than $C$, thus

$$\lim_{\lambda \to \infty} U = 3C. \quad (30)$$

On the other hand, the Hund parameter can be approximated with $J \approx C$ because $B$ is order of magnitude smaller than $C$ for $\lambda > 0$ (see Fig. 2). For $\lambda \to \infty$, $B$ actually goes to zero faster than $C$ (see Eq. 26). Thus in the strong screening limit, Hund parameters $J_{\alpha \neq \beta}$ in Table. 1 approaches same values regardless of the combination

$$\lim_{\lambda \to \infty} J_{\alpha \neq \beta} = C. \quad (31)$$

## Table 2

$\lambda$ values for various Transition Metal elements (Ion), extracted from $U$ values of Density Functional Theory (DFT) studies. $\zeta$ values are taken from Ref. [18]. (See B)

| $Z$ | Element | $\lambda$ (Å$^{-1}$) | Refs. |
|---|---|---|---|
| 22 | Ti | 1.32 ± 0.43 | TiO$_2$[20, 21, 22] |
| 23 | V | 1.50 ± 1.11 | VO[23, 24],V$_2$O$_3$[24],VO$_2$[20, 24],V$_2$O$_5$[24] |
| 24 | Cr | 1.62 ± 0.41 | Cr$_2$O$_3$[21, 23, 24], CrO$_3$[24] |
| 25 | Mn | 1.59 ± 0.56 | MnO[25, 21, 23, 24, 26, 27], Mn$_2$O$_3$[24], MnO$_2$[24, 28] |
| 26 | Fe | 1.82 ± 0.47 | FeO[25, 21, 23, 24, 27, 26],Fe$_2$O$_3$[21, 24, 26] |
| 27 | Co | 1.88 ± 0.71 | CoO[21, 25, 23, 24, 27] |
| 28 | Ni | 1.95 ± 0.83 | NiO[21, 25, 26, 29, 23, 27] |
| 29 | Cu | 1.97 ± 0.40 | CuO[23, 30, 31], Cu$_2$O[21, 30, 26] |
| 30 | Zn | 2.03 ± 0.66 | ZnO[32, 33, 34, 35] |

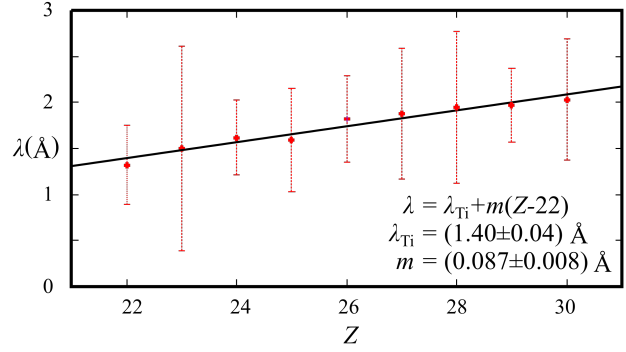

**Figure 4:** [Color online] Screening constant $\lambda$ for transition metals. The screening constant increases as the atomic number $Z$ increases. $Z$ =22, 23, 24, 25, 26, 27, 28, 29 and 30 are Ti, V, Cr, Mn, Fe, Co, Ni, Cu, and Zn, respectively.

These expressions of $U$ and $J$ can be used for estimating their realistic values from the values of the shielding constant $\zeta$ and the screening constant $\lambda$. The former is the property of the orbital wave function that is well studied in atomic physics [17, 18, 19]. On the other hand, the screening constant $\lambda$ depends on the screening interaction. As far as we know, the value of $\lambda$ for transition metal has not been well-described in the previous researches. Table 2 summarizes the values of $\lambda$ for transition metals from previous DFT studies.

Comparing the values of screening constant for varius combinations of transition-metal oxides, we find that the screening constant increases linearly from Ti to Zn, as shown in Fig. 4

$$\lambda = \lambda_{Ti} + m(Z - 22) \qquad (32)$$

where $\lambda_{Ti} = (1.40 \pm 0.04)$ Å and the gradient $m = (0.087 \pm 0.008)$ Å. By using Eq. 32 with the value of $\zeta$, one can predict a realistic value of $U$ and $J$ for modeling the material with the generalized Hubbard model.

## 5. Conclusions and remarks

Material modeling that uses generalized Hubbard model requires Hubbard's $U$ an Hund's $J$ parameters. Although the analytic value of $U$ and $J$ based on bare Coulomb potential are described by Racah's $A, B, C$ parameters, the values are too large for realistic transition-metal-based materials.

By replacing the bare Coulomb potential with Yukawa-type screened Coulomb potential, we show that $A, B, C$ parameters for screened Coulomb potential with screening constant $\lambda$ can be obtained by replacing $r_<^l/r_>^{l+1}$ with

$$f_l(r_1, r_2) = (2l+1)\lambda i_l(\lambda r_<) k_l(\lambda r_>),$$

## Table 3

Modified Spherical Bessel functions $i_n, k_n$ for $n = 0, 2, 4$ [15, 16]

| $n$ | $i_n(x)$ |
|---|---|
| 0 | $\frac{\sinh x}{x}$ |
| 2 | $\left(1+\frac{3}{x^2}\right)\frac{\sinh x}{x} - \frac{3\cosh x}{x^2}$ |
| 4 | $\left(1+\frac{45}{x^2}+\frac{105}{x^4}\right)\frac{\sinh x}{x} - \left(10+\frac{105}{x^2}\right)\frac{\cosh x}{x^2}$ |

| $n$ | $k_n(x)$ |
|---|---|
| 0 | $\frac{1}{x}e^{-x}$ |
| 2 | $\left(\frac{1}{x}+\frac{3}{x^2}+\frac{3}{x^3}\right)e^{-x}$, |
| 4 | $\left(\frac{1}{x}+\frac{10}{x^2}+\frac{45}{x^3}+\frac{105}{x^4}+\frac{105}{x^5}\right)e^{-x}$ |

where $i_n$ and $k_n$ are the modified spherical Bessel functions of the first kind and second kind, respectively. $\lambda$ is the screening constant that characterized the screened Coulomb interaction. We used the corresponding modified Slater integrals to determine the modified $A, B, C$ parameters and compared to the case of the bare Coulomb potential.

The $A, B, C$ parameters determine the Hubbards's $U$ and Hund's $J$ parameters. We discuss the cases for weak and strong screenings, and describe the asymptotic behavior. While both $U$ and $J$ decrease as $\lambda$ increases, $U$ decreases faster than $J$. This is because $U$ is affected by first order correction of $\lambda$, while $J$ is affected by the second-order.

The realistic $U$ and $J$ values can be determined from the values of $\zeta$, which value is well studied in atomic physics, and the screening constant $\lambda$ of the screened Coulomb interaction. We found that the $\lambda$ for transition metal increases as the atomic number increase from Ti to Zn. The expression for realistic values of $J$ and $U$ should be able to help for theoretical modeling of materials based on the generalized Hubbard model.

While we focus on the transition metal, the expression of A,B,C parameters can be applied to more general cases. The weak screening case $\lambda \ll \zeta$, in particular, should be useful for rare-earth-based materials, because rare earth ions have large $\zeta$ values.

## CRediT authorship contribution statement

**Adam B. Cahaya**: Conceptualization, Methodology, Formal analysis, Investigation, Writing - original draft, Writing - review & editing. **Anugrah Azhar**: Data curation, Investigation, Resources. **Muhammad Aziz Majidi**: Conceptualization, Supervision, Funding acquisition.

## Declaration of competing interest

The authors declare that they have no known competing financial interests or personal relationships that could have appeared to influence the work reported in this paper.

## A. Modified Spherical Bessel functions of the first and second kinds

Modified Spherical Bessel functions are related to modified Bessel functions the first and second kind $I_n, K_n$ as[36]

$$i_n(x) = \sqrt{\frac{\pi}{2x}} I_{n+\frac{1}{2}}(x), \qquad (33)$$

$$k_n(x) = \sqrt{\frac{2}{\pi x}} K_{n+\frac{1}{2}}(x). \tag{34}$$

They are also related to spherical Bessel functions of the first and second kind $j_n, y_n$ as[36]

$$i_n(x) = i^{-n} j_n(ix), \tag{35}$$
$$k_n(x) = -i^n \left(j_n(ix) + i y_n(ix)\right). \tag{36}$$

The explisit forms for $n = 0, 2, 4$ are listed in Table 3.

The asymptotic behaviors of the modified spherical Bessel functions are [37, 36]

$$i_n(x) \simeq \begin{cases} \frac{x^n}{1 \cdot 3 \cdot 5 \cdots (2n+1)}\left(1 + \frac{x^2}{2(2n+3)}\right), & x \ll 1, \\ \frac{e^x}{2x}\left(1 - \frac{n(n+1)}{2x}\right), & x \gg 1, \end{cases} \tag{37}$$

$$k_n(x) \simeq \begin{cases} (-1)^n i_l(x) + \frac{1 \cdot 3 \cdots (2n-1)}{x^{n+1}}\left(1 - \frac{x^2}{2(2n-1)}\right), & x \ll 1, \\ \frac{e^{-x}}{x}\left(1 + \frac{n(n+1)}{2x}\right), & x \gg 1. \end{cases} \tag{38}$$

This asymptotic functions is useful for discussion of weak- and strong-screening limit cases in Sec. 3.2.

## B. Data set of $\lambda$ values

See Table 4

**Table 4**
$U$ and $\lambda$ values for various transition metal ions

| Ion | $\zeta$ (Å$^{-1}$) | $U$ (eV) | $\lambda$ (Å$^{-1}$) |
|---|---|---|---|
| TiO$_2$ (Ti$^{4+}$) [20] | 2.89 | 2.30 | 1.82 |
| TiO$_2$ (Ti$^{4+}$) [21] | 2.89 | 3.00 | 1.42 |
| TiO$_2$ (Ti$^{4+}$) [22] | 2.89 | 3.40 | 1.25 |
| TiO$_2$ (Ti$^{4+}$) [22] | 2.89 | 5.00 | 0.78 |
| VO (V$^{2+}$) [23] | 2.96 | 3.10 | 1.44 |
| VO (V$^{2+}$) [24] | 2.96 | 4.12 | 1.06 |
| V$_2$O$_3$ (V$^{3+}$) [24] | 3.05 | 4.99 | 0.89 |
| VO$_2$ (V$^{4+}$) [20] | 3.14 | 1.10 | 3.74 |
| VO$_2$ (V$^{4+}$) [24] | 3.14 | 5.14 | 0.91 |
| V$_2$O$_5$ (V$^{5+}$) [24] | 3.23 | 5.12 | 0.98 |
| Cr$_2$O$_3$ (Cr$^{3+}$) [24] | 3.29 | 2.73 | 1.99 |
| Cr$_2$O$_3$ (Cr$^{3+}$) [21] | 3.29 | 3.00 | 1.83 |
| Cr$_2$O$_3$ (Cr$^{3+}$) [23] | 3.29 | 3.50 | 1.58 |
| CrO$_3$ (Cr$^{6+}$) [24] | 3.29 | 4.99 | 1.06 |
| MnO (Mn$^{2+}$) [25] | 3.44 | 2.00 | 2.82 |
| MnO (Mn$^{2+}$) [21] | 3.44 | 3.00 | 2.00 |
| MnO (Mn$^{2+}$) [23] | 3.44 | 4.00 | 1.51 |
| MnO (Mn$^{2+}$) [24] | 3.44 | 4.94 | 1.19 |
| MnO (Mn$^{2+}$) [26] | 3.44 | 5.00 | 1.17 |
| MnO (Mn$^{2+}$) [27] | 3.44 | 6.00 | 0.92 |
| Mn$_2$O$_3$ (Mn$^{3+}$) [24] | 3.53 | 4.05 | 1.57 |
| MnO$_2$ (Mn$^{4+}$) [28] | 3.62 | 3.90 | 1.71 |
| MnO$_2$ (Mn$^{4+}$) [24] | 3.62 | 4.78 | 1.38 |
| FeO (Fe$^{2+}$) [21, 25] | 3.69 | 3.00 | 2.27 |
| FeO (Fe$^{2+}$) [26] | 3.69 | 3.70 | 1.87 |
| FeO (Fe$^{2+}$) [23] | 3.69 | 4.00 | 1.73 |
| FeO (Fe$^{2+}$) [24] | 3.69 | 4.10 | 1.69 |
| FeO (Fe$^{2+}$) [27] | 3.69 | 7.00 | 0.87 |
| Fe$_2$O$_3$ (Fe$^{3+}$) [21] | 3.78 | 3.00 | 2.38 |
| Fe$_2$O$_3$ (Fe$^{3+}$) [24] | 3.78 | 3.47 | 2.08 |
| Fe$_2$O$_3$ (Fe$^{3+}$)[26] | 3.78 | 4.30 | 1.69 |
| CoO (Co$^{2+}$) [21, 25] | 3.93 | 3.00 | 2.56 |
| CoO (Co$^{2+}$) [23] | 3.93 | 3.30 | 2.35 |
| CoO (Co$^{2+}$) [24] | 3.93 | 4.89 | 1.60 |
| CoO (Co$^{2+}$) [27] | 3.93 | 7.00 | 1.02 |
| NiO (Ni$^{2+}$) [21] | 4.18 | 3.00 | 2.86 |
| NiO (Ni$^{2+}$) [25] | 4.18 | 3.00 | 2.86 |
| NiO (Ni$^{2+}$) [26] | 4.18 | 3.80 | 2.32 |
| NiO (Ni$^{2+}$) [29] | 4.18 | 6.20 | 1.38 |
| NiO (Ni$^{2+}$) [23] | 4.18 | 6.40 | 1.33 |
| NiO (Ni$^{2+}$) [27] | 4.18 | 8.00 | 0.97 |
| CuO (Cu$^{2+}$) [23] | 4.42 | 4.00 | 2.47 |
| CuO (Cu$^{2+}$) [30] | 4.42 | 5.00 | 1.99 |
| CuO (Cu$^{2+}$) [31] | 4.42 | 7.14 | 1.31 |
| Cu$_2$O (Cu$^{4+}$) [21] | 4.60 | 5.00 | 2.15 |
| Cu$_2$O (Cu$^{4+}$) [30] | 4.60 | 5.00 | 2.15 |
| Cu$_2$O (Cu$^{4+}$) [26] | 4.60 | 6.00 | 1.77 |
| ZnO (Zn$^{2+}$) [32] | 4.66 | 4.00 | 2.74 |
| ZnO (Zn$^{2+}$) [33] | 4.66 | 4.70 | 2.35 |
| ZnO (Zn$^{2+}$) [34] | 4.66 | 5.00 | 2.21 |
| ZnO (Zn$^{2+}$) [32] | 4.66 | 6.00 | 1.82 |
| ZnO (Zn$^{2+}$) [35] | 4.66 | 9.30 | 1.01 |